\documentclass[sigconf]{acmart}

\AtBeginDocument{%
  \providecommand\BibTeX{{%
    \normalfont B\kern-0.5em{\scshape i\kern-0.25em b}\kern-0.8em\TeX}}}

\copyrightyear{2023}
\acmYear{2023}
\setcopyright{acmlicensed}\acmConference[WWW '23 Companion]{Companion Proceedings of the ACM Web Conference 2023}{April 30-May 4, 2023}{Austin, TX, USA}
\acmBooktitle{Companion Proceedings of the ACM Web Conference 2023 (WWW '23 Companion), April 30-May 4, 2023, Austin, TX, USA}
\acmPrice{15.00}
\acmDOI{10.1145/3543873.3587641}
\acmISBN{978-1-4503-9419-2/23/04}

\begin{document}

\title{Cultural Differences in Signed Ego Networks on Twitter: An Investigatory Analysis}

\author{Jack Tacchi}
\affiliation{%
  \institution{IIT-CNR}
  \city{Pisa}
  \country{Italy}
}
\email{jack.tacchi@sns.it}

\author{Chiara Boldrini}
\affiliation{%
  \institution{IIT-CNR}
  \city{Pisa}
  \country{Italy}
}
\email{chiara.boldrini@iit.cnr.it}

\author{Andrea Passarella}
\affiliation{%
  \institution{IIT-CNR}
  \city{Pisa}
  \country{Italy}
}
\email{andrea.passarella@iit.cnr.it}

\author{Marco Conti}
\affiliation{%
  \institution{IIT-CNR}
  \city{Pisa}
  \country{Italy}
}
\email{marco.conti@iit.cnr.it}


\begin{abstract}
  \vspace{-2pt}
  Human social behaviour has been observed to adhere to certain structures. One such structure, the Ego Network Model (ENM), has been found almost ubiquitously in human society. Recently, this model has been extended to include signed connections. While the unsigned ENM has been rigorously observed for decades, the signed version is still somewhat novel and lacks the same breadth of observation. Therefore, the main aim of this paper is to examine this signed structure across various categories of individuals from a swathe of culturally distinct regions. Minor differences in the distribution of signs across the SENM can be observed between cultures. However, these can be overwhelmed when the network is centred around a specific topic. Indeed, users who are engaged with specific themes display higher levels of negativity in their networks. This effect is further supported by a significant negative correlation between the number of "general" topics discussed in a network and that network's percentage of negative connections. These findings suggest that the negativity of communications and relationships on Twitter are very dependent on the topics being discussed and, furthermore, these relationships are more likely to be negative when they are based around a specific topic.
  \vspace{-5pt}
\end{abstract}


\vspace{-10pt}


\maketitle

\section{Introduction}
\label{sec:introduction}

\vspace{-2pt}

Humans communicate on a scale and with a complexity far beyond that of any other animal. Indeed, it has even been hypothesised that the main reason humans evolved larger brains was to be able to maintain larger groups that drastically improved our ability to survive and reproduce~\cite{Dunbar_1998}. In fact, the size of a species' brain (specifically its neocortex) is strongly correlated to the number of relationships it can maintain~\cite{Dunbar_1992}. For humans, this number is around 150 (known as Dunbar's number) and has been observed in a variety of contexts: from Neolithic village populations in Mesopotamia~\cite{Oates_1977} to the sizes of independent units in professional armies of the 16\textsuperscript{th} to 20\textsuperscript{th} centuries~\cite{Dunbar_1993}, and even in modern-day communities in online social media~\cite{Arnaboldi_2015}.

\begin{figure}[h]
  \centering
  \includegraphics[scale=0.27]{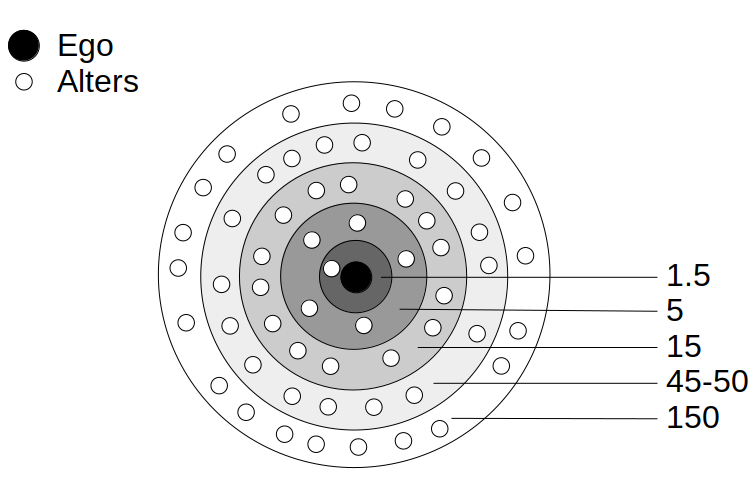}
  \caption{The Ego Network Model, with the expected sizes of each circle (for humans).}
  \Description{A graphical representation of the Ego Network Model. The Ego, represented as a black dot, sits at the centre. It is surrounded by 5 grey circles of increasingly larger size and lighter colour. All of these circles are centred around the Ego. The circles are labelled according to the number of Alters they are expected (on average) to contain, the smallest being 1.5, followed by 5, 15, 45 to 50 and then 150.}
  \label{fig:ego_network_model}
  \vspace{-25pt}
\end{figure}

What's more, by taking the viewpoint of a single individual in a social network, further patterns emerge within these groups of 150. Most notably, when the relationships of the selected individual are organised based on their strength, they can usually be represented as a series of concentric circles of increasing size but decreasing relationship strength. This representation is known as the Ego Network Model (ENM)~\cite{Dunbar_1995}, with the individual at the centre being referred to as the Ego and their contacts being referred to as Alters. An illustration of the ENM can be seen in Figure~\ref{fig:ego_network_model}. As the circles of the ENM also stem from the innate cognitive limits of the neocortex, they too have regular and predictable sizes; these being, in order of increasing size but decreasing tie strength to the Ego: 5 (support clique), 15 (sympathy group), 45-50 (affinity group) and 150 (active network). There have also been observations of an additional group with a mean size of 1.5 Alters at the very centre~\cite{Arnaboldi_2013a}. Each circle is larger than the one before it by a factor of roughly 3. This ratio of 3 has been observed not only in humans but also in other primates and even certain species of bird (although the group sizes themselves are smaller for other animals)~\cite{Dunbar_2014}.

Of course, as the Alters of the ENM are grouped based on the strength of their connection to the Ego, how their strength is measured is extremely important. Traditionally, this has been done by measuring contact frequency, a clearly defined and easy-to-measure metric that has been shown to be a good proxy for relationship strength~\cite{Gilbert_2009}. However, relationships are very complex things and, by merely measuring the contact frequency, a lot of the more qualitative information is being lost. 

With regard to network research in general, adding the additional dimension of signed connections (i.e. positive or negative) often reveals huge amounts of additional information about the surrounding network. This is particularly true for negative connections, which can provide a disproportionately large amount of information in certain contexts~\cite{Leskovec_2010}. Because of this, signs have often been used to great benefit for many tasks within this field, such as community detection~\cite{Esmailian_2015} and information diffusion~\cite{Shi_2016}. In addition to this, online social networks provide a plethora of text-based communications that can easily be converted into signs via sentiment analysis. Therefore, the incorporation of signed connections seems to be a natural extension of the ENM that could help to retain some of the more qualitative aspects of the data that are being lost due to only measuring contact frequency. Indeed, a recent work did just this using data from Twitter, coining the Signed Ego Network Model (SENM)~\cite{Tacchi_2022}. 

Given the ubiquitousness of the ENM in human societies, the question naturally arises as to whether the pattern of signs in the SENM is as equally widespread. Therefore, the main aim of this paper is to observe the SENMs of individuals from a variety of locations around the world to see whether the SENM can be generalised as much as its unsigned counterpart and what additional information can be obtained via the inclusion of signed connections. What's more, these observations were done using multilingual, rather than English-only, data. This culminated in 3 main findings. First, although there are some minor differences in the negativity of signed Ego Networks for generic users from different cultures, these differences can be overwhelmed when the network is centred around a specific topic. Second, networks that are "themed" around a specific topic are consistently more negative than generic networks, both overall and at each level of the SENM. Lastly, of the different topics observed, "general" (meaning no specific topic, see Subsection~\ref{subsec:most_popular_topics}) is the only topic to have a significant (negative) correlation with the percentage of negative relationships in a network. These findings suggest that the negativity of communications and relationships on Twitter is very dependent on the topics being discussed and that, when they are based on a specific topic, these relationships are more likely to be negative.

\section{Background}

\subsection{Positive and Negative Relationships}
\label{subsec:positive_and_negative_relationships}
Network research is not the only scientific area that has investigated the signs of relationships. Findings from psychology have revealed that obtaining even just the signs of individual interactions generated by a relationship can tell us a lot about that relationship as a whole. For instance, a ratio of 5 positive interactions to each negative interaction (or roughly 17\%) appears to be a significant threshold for many different types of relationships. Relationships that produce more negative interactions than this are significantly less likely to continue and are more likely to cause negative afflictions on the people involved. For instance, marriages with too many negative exchanges are more likely to end in divorce~\cite{Gottman_1995} and overly negative parental relationships can increase the probability of the child developing behavioural problems and struggling in school~\cite{Hart_1995}. 

In previous works, the number of negative relationships observed in a network has varied slightly from context to context but is usually within the range of 12.3\% to 22.6\%~\cite{Leskovec_2010,Tang_2015}. These observations were made using data from Epinions (a general consumer review site), Slashdot (a technology news site) and Wikipedia (an online encyclopedia). However, these relationships were explicitly signed and publicly visible to the other individuals in the networks. Therefore, they may have been influenced by an effect known as social capital, whereby individuals who have many relationships in common (i.e. close-knit groups) feel strong social pressure to get along and maintain positive relations with the surrounding community~\cite{Coleman_1988}. Indeed, similar relationships from Twitter, where the signs of relationships are not explicitly visible have displayed higher percentages of negative relationships: from 21.83\% to as high as 54.89\%~\cite{Tacchi_2022}. Given that Twitter is also the source of data for this paper (described further in Subsection~\ref{subsec:datasource}), it would be expected to find negative percentages around the same range. 
\vspace{-5pt}

\subsection{Signed Ego Network Model}
\label{subsec:signed_ego_network_model}
By combining signed connections with the ENM, the SENM has revealed some hitherto obscured information about the distribution of negative relationships in Ego Networks. Similar to analyses of the unsigned ENM, particular attention is paid to the mean number of signs in the overall networks, as well as for each circle. Surprisingly, the inner circles, which are considered to be an Ego's closest and most trusted connections, were found to be the most negative~\cite{Tacchi_2022}. What's more, this was especially true for users who are more engaged with the platform and, in particular, journalists, who use the platform for professional reasons. The number of negative relationships also varied a lot more across the circles for the journalists and some other types of specialised users compared to regular (or generic) users, who use the platform for non-professional reasons. These findings were further supported by the fact that the active networks (i.e. up to around 150 Alters) were more negative overall than the full networks. It could therefore be interesting to investigate whether this is an effect that is particular to journalists, or if other types of overly engaged users will display similar levels of negativity.

\section{Methodology and Data}

\subsection{Computation of Signed Ego Networks}
\label{subsec:computation_of_signed_ego_networks}
As mentioned in Section~\ref{sec:introduction}, the ENM depicts the network from the viewpoint of a single Ego, with their contacts arranged around them in a series of concentric circles. While the number of close relationships tends to be close to 150 for humans, people will certainly make more than 150 connections across their lifetime. Indeed, the importance of each contact will change greatly over time, potentially moving between the circles or even leaving that person's active network~\cite{Boldrini_2018}. An Alter is said to be outside an Ego's active network when they are contacted less than once a year~\cite{Hill_2003}, whereupon the cognitive effort put into maintaining that relationship will be negligible, and so it is standard practice to consider only the active network when analysing Ego Networks. When analysing an Ego Network, researchers will usually pay particular attention to the mean size of the active network, as well as those of the individual circles. 

In order to obtain the Signed Ego Networks, the unsigned versions are first computed. This is done by clustering the Alters in each Ego's active network based on their frequency of contact with the Ego. Various clustering algorithms have previously been used for this purpose; however, MeanShift~\cite{Fukunaga_1975} was chosen for the current work. One significant advantage of MeanShift is that it automatically determines the optimum number of clusters, which can help to better consider some of the slight differences between individuals that have been observed in Ego Networks in online contexts~\cite{Boldrini_2018}. These discrepancies are largely due to the varying degrees of time and cognitive effort that individuals spend online (i.e. the extent to which the online part of their Ego Network has developed).

Next, a "positive" or "negative" sign is obtained for each relationship in the Ego Networks. This is done by first gathering all text-based interaction Tweets (Replies, Mentions and Quote Retweets) from each Ego to each of their Alters. Then a sentiment analysis model (see Subsection~\ref{subsec:multilingual_model}) is used to label each individual interaction as either "positive", "neutral" or "negative". Finally, an overall label is assigned based on the frequency of "negative" interactions. This is done using the threshold interaction ratio of 17\% mentioned in Section~\ref{subsec:positive_and_negative_relationships}. Relationships with more negative interactions than this are considered "negative" and those with equal to or fewer are considered "positive". 

Once both the unsigned Ego Networks and the relationship signs have been computed, the Signed Ego Networks can be easily obtained by matching the signs of each relationship to those of the corresponding Ego-Alter pair.
\vspace{-7pt}

\subsection{Multilingual Sentiment Model}
\label{subsec:multilingual_model}
The model used for signing the individual interactions is an important consideration as it is the foundation of the relationship labels and, therefore, the signs of the SENM. The initial paper on the SENM~\cite{Tacchi_2022} used a monolinguistic model called Valence Aware Dictionary and sEntiment Reasoner (VADER)~\cite{Hutto_2014}. Unfortunately, given that the datasets here contain a variety of languages, it is not possible to use the same model. One alternate method could be to choose a similar model for each language. However, given that some languages have received vastly more research and resources than others, the performances of these models would likely vary significantly. Furthermore, as sentiment models for different languages cannot be tested on a common dataset, it can be difficult to compare them reliably. Therefore, a single, multilingual model was chosen for this work. 

The chosen model is XLM-T~\cite{Barbieri_2022}\footnote{Documentation available at: \url{https://huggingface.co/cardiffnlp/twitter-xlm-roberta-base-sentiment}.}. This model is based on a multilingual, transformer-based model, XLM-R~\cite{Conneau_2019}, but was additionally trained on 198 million Tweets from over 60 languages. For the datasets used in the current paper (described in Subsection~\ref{subsec:datasets}), the most commonly occurring languages are English (66.47\%), Dutch (6.96\%), Spanish (5.54\%), Portuguese (4.65\%), German (2.83\%), Italian (2.58\%) and French (1.41\%)\footnote{The percentages correspond to the proportions of interactions in each language across all the datasets, before any preprocessing. Also, it is important to note that although English appears as the vast majority, this is because of the much larger size of the Baseline dataset. Other than the Baseline, only West Africa was majority English.}, all of which are in the top 20 most frequent languages used to train XLM-T. This model was tested for the task of sentiment analysis on a multilingual corpus consisting of 8 equally-represented languages. It achieved F1 scores of between 67.91 and 77.35 for 6 of the top 7 languages for the current study (Dutch being the missing language). So, although this model doesn't completely resolve the problem of different performances for different languages, it does display a fairly standard level of performance across all of the languages that it has been tested on and that are relevant to this paper.

\subsection{Data}

\subsubsection{Data Source}
\label{subsec:datasource}

All of the datasets in this paper have been collected from Twitter, which has long been a reliable source of data for Ego Network research (for example~\cite{Arnaboldi_2013a,Boldrini_2018}). This is primarily due to its vast userbase, generally public data and easily accessible API. Furthermore, Twitter users are able to directly communicate with one another in a few, easily identifiable ways: Replies, Mentions and Retweets. These interactions make it very easy to assign an Ego and an Alter (or Alters) to each communication and, therefore, to know when any cognitive effort has been spent on maintaining a relationship.

The data collection was conducted using two key endpoints of the Twitter API: Twitter Search and User Tweet Timelines. The former provides a stream of Tweets based on a provided search query in (reverse) chronological order, similar to manually searching a query on the Twitter homepage. This is useful for finding tweets on specific topics as well as identifying users who are engaged with said topics. A User Tweet Timeline is a collection of all the public tweets created by a given user. This includes all the aforementioned communication Tweets, as well as standard, non-communicative Tweets, which are irrelevant here (because the ENM focuses on the interactions between peers).

All the datasets used in this paper consist of a series of User Tweet Timelines, each Timeline corresponding to a single Ego.

\subsubsection{Datasets}
\label{subsec:datasets}

The first dataset that was collected was a baseline dataset\footnote{The Tweet IDs for this dataset are provided at \url{https://zenodo.org/record/7717006}.}. This dataset was collected using a snowball sampling methodology, which first gathered the User Tweet Timelines of a small set of initial seed users, followed by those of these users' Alters, and then those of their Alters' Alters. The initial seeds were a set of 31 randomly selected users taken from an extremely large, preexisting dataset of Ego Networks (this original dataset is itself a snowball dataset with the initial seed being Barack Obama)~\cite{Arnaboldi_2013a}, with the only prerequisite being that at least one of their alters had to have already been selected as a seed user (with the obvious exception of the very first user). The original dataset was collected in November 2012; therefore, the Timelines of the selected seeds had to be recollected to ensure their Alters were up to date. The baseline dataset was collected between the 27\textsuperscript{th} April and the 25\textsuperscript{th} May 2022.

The next set of datasets was collected using the same snowball sampling method. This time, however, the initial users were selected based on their locations. As the number of active Twitter users varies enormously from country to country, it is not possible to collect meaningfully large datasets for every region of the globe. Because of this, countries were prioritised if they had large active userbases. To aid in this choice, the locations in the baseline dataset were examined (an overview of the number of users for the top locations can be seen in Appendix~\ref{app:locations}). This led to the selection of 4 culturally-distinct parts of the world where data were reasonably available: the Mediterranean (Spain, France, Italy, Greece), Northern Europe (Germany, Netherlands, Sweden), West Africa (Nigeria, Senegal, Ghana) and South America (Brazil, Colombia, Venezuela)\footnote{The Tweet IDs for these datasets are provided at \url{https://zenodo.org/record/7717047}.}.

However, there are a few problems when it comes to collecting users based on their location. Namely, users are free to put any location they wish on their Twitter profile, even if that location is incorrect or fictitious. For this reason, extra attention was paid to the selection of the regional datasets' initial seed users, 3 of which were collected for each country by querying Twitter Search for the name of each country in its most widely spoken official language. The profiles of these users were then manually checked to ensure that they were generic users from the desired country. Of course, it was not possible to manually check every user that was collected, so this was only done for the initial seeds. Unfortunately, the highly differing amounts of data available in different locations further exacerbate the aforementioned problem of accurately collecting enough data for each country. Consequently, the countries were grouped together, based primarily on geographical proximity and cultural similarity, but also on the availability of data (which is why, for example, France is placed in the Mediterranean dataset, despite not being entirely Mediterranean). These datasets were collected between the 16\textsuperscript{th} June 2022 and the 26\textsuperscript{th} July 2022.

Because the users in these regional datasets, as well as those of the baseline, were not selected based on any common interest or profession, they are considered to be predominantly "generic" users. Generic users, among other things, are likely to be less engaged with the Twitter platform when compared to other types of users. So, the 5 generic datasets used in this paper are:

\begin{itemize}
    \item Baseline
    \item Mediterranean (Spain, France, Italy, Greece)
    \item Northern Europe (Germany, Netherlands, Sweden)
    \item West Africa (Nigeria, Senegal, Ghana)
    \item South America (Brazil, Colombia, Venezuela)
\end{itemize}

The next set of datasets was taken from previous work investigating the Ego Networks of journalists from around the world~\cite{Toprak_2021}. It has previously been observed that, for English-speaking users, the Signed Ego Networks of journalists are more negative than those of generic users~\cite{Tacchi_2022}. Therefore, journalist data were selected that correspond to 3 of the 4 aforementioned regions (there was no corresponding dataset for West Africa), so as to make similar comparisons between non-English speaking generic and specialised users. These non-English speaking journalists datasets are: Italian Journalists, Dutch Journalists and Brazilian Journalists. They were originally collected between the 14\textsuperscript{th} and the 17\textsuperscript{th} January 2018 from a list of verified Twitter journalist accounts. 

Given that the previously discussed differences between the Ego Networks of generic users and journalists are expected to be due to differing levels of engagement with Twitter~\cite{Boldrini_2018}, a further 3 datasets were collected with the aim of obtaining highly-engaged non-journalists to further investigate this effect. The users in these datasets were identified using Twitter Search. However, the choice of search query was somewhat problematic; most popular topics are covered by journalists to some degree, and topics that are not popular are unlikely to provide enough data for meaningful results. On top of this, the topic needs to be specific to users from a certain region. One solution for this is to use query terms related to reality TV shows, as these shows tend to have large, highly-engaged audiences, yet receive relatively small amounts of press coverage. The datasets (and their search terms) are Italian Reality TV (\#XF2022, \#GFVIP), Dutch Reality TV (\#HollandsGotTalent, \#IkVertrek), Brazilian Reality TV (\#XFactorBR, \#BBB22)\footnote{The Tweet IDs for these datasets are provided at \url{https://zenodo.org/record/7716860}.}. The collection of these data was performed between the 21\textsuperscript{st} and the 29\textsuperscript{th} January 2023.

For Italy and Brazil, the local versions of "The X Factor" and "Big Brother" were used. For the Netherlands, these shows have the same titles as their British and American counterparts. So, to avoid collecting users from other countries, similar, non-linguistically ambiguous TV shows were chosen: "Holland's Got Talent" and "Ik Vertrek". The Tweet Timelines of users who Tweeted using one or more of the search terms were collected. Although these users could spend the majority of their time tweeting about other topics, a look at the most common topics discussed by these users revealed that this does not seem to be the case. Indeed, most of their communications are discussing topics related to reality TV (see Subsection~\ref{subsec:most_popular_topics}). 

As the journalist and reality TV datasets were collected based on specific themes, they are referred to collectively as the "themed" datasets. Based on the findings previously discussed, combined with the fact that these users are more likely to be part of an online community focused around their specific topics, it is expected that they will be more engaged with the Twitter platform than the generic users. The 6 themed datasets used in this paper are:

\begin{itemize}
    \item Italian Journalists
    \item Dutch Journalists
    \item Brazilian Journalists
    \item Italian Reality TV (\#XF2022, \#GFVIP)
    \item Dutch Reality TV (\#HollandsGotTalent, \#IkVertrek)
    \item Brazilian Reality TV (\#XFactorBR, \#BBB22)
\end{itemize}

\subsubsection{Preprocessing}
\label{subsec:preprocessing}
Because the ENM's structure is a direct result of an individual's cognitive limits, it is important to ensure that all Egos included in the data are individual humans, i.e. not bots or teams of humans working together. In order to remove these undesired types of users, a Support Vector Machine (SVM)~\cite{Cortes_1995} was trained on a sample of 500 users that had been manually annotated as either "people" or "others". Both the choice of classifier and the training data have previously been established in ENM research~\cite{Arnaboldi_2013a}, where they obtained a mean accuracy of 81.3\% and a mean false positive rate of 8.3\% using k-fold cross-validation (with k=5). Any Egos labelled as "others" by this classifier were removed\footnote{This was not necessary for the journalist datasets as these were collected from a list of verified individuals.}.

Next, Egos who were not very active on Twitter, and who were therefore unlikely to have substantially developed an Ego Network on the platform, were removed. This was done for Egos whose Timeline consisted of fewer than 2,000 Tweets, whose Timeline lasted less than 6 months or who tweeted less than once every 3 days for more than 50\% of the months they were active. Finally, Alters were removed if their Ego interacted with them less than once a year. These parameters are based on pre-existing standards within the ENM field~\cite{Arnaboldi_2015,Boldrini_2018} and stem from psychological and anthropological research~\cite{Hill_2003}.

After removing the bots and groups, the remaining Ego Networks correspond to the full network of each Ego; after removing the inactive Egos and Alters, they correspond to the active part, as mentioned in Section~\ref{sec:introduction}. Descriptive statistics of each dataset before and after preprocessing can be seen in Table~\ref{tab:descriptives_full_network} and Table~\ref{tab:descriptives_active_network} respectively.

\begin{table}
  \caption{Number of Egos, Relationships and Interactions in the full Ego Networks}
  \label{tab:descriptives_full_network}
  \begin{tabular}{lrrr}
    \toprule
    Dataset & Egos & Relationships & Interactions \\
    \midrule
    Baseline & 12,200 & 9,636,070 & 77,088,560\\
    Mediterranean & 2,529 & 2,047,634 & 8,338,571\\
    Northern Europe & 1,504 & 1,309,802 & 4,883,393\\
    West Africa & 1,669 & 1,285,608 & 4,656,912\\
    South America & 1,203 & 917,741 & 3,648,586\\
    \bottomrule
    Italian Journalists & 485 & 381,169 & 1,426,780\\
    Dutch Journalists & 4,298 & 2,434,647 & 8,646,733\\
    Brazilian Journalists & 890 & 596,879 & 2,084,906\\
    Italian Reality TV & 1,039 & 708,193 & 2,379,952\\
    Dutch Reality TV & 1,139 & 734,936 & 2,727,800\\
    Brazilian Reality TV & 995 & 689,096 & 2,230,007\\
    \bottomrule
  \end{tabular}
  \vspace{-5pt}
\end{table}

\begin{table}
  \caption{Number of Egos, Relationships and Interactions in the active Ego Networks}
  \label{tab:descriptives_active_network}
  \begin{tabular}{lrrr}
    \toprule
    Dataset & Egos & Relationships & Interactions \\
    \midrule
    Baseline & 4,049 & 574,585 & 8,593,290\\
    Mediterranean & 878 & 120,068 & 2,191,666\\
    Northern Europe & 552 & 82,237 & 1,273,881\\
    West Africa & 396 & 55,884 & 884,321\\
    South America & 217 & 25,205 & 441,158\\
    \bottomrule
    Italian Journalists & 203 & 30,409 & 489,008\\
    Dutch Journalists & 1,316 & 179,668 & 2,702,275\\
    Brazilian Journalists & 154 & 20,348 & 278,631\\
    Italian Reality TV & 160 & 18,884 & 291,213\\
    Dutch Reality TV & 230 & 24,082 & 441,694\\
    Brazilian Reality TV & 154 & 15,685 & 234,734\\
    \bottomrule
  \end{tabular}
\end{table}

\section{Results}

\subsection{Preliminaries: VADER vs XLM-T}
As discussed in Section~\ref{subsec:multilingual_model}, the switch was made from the monolingual VADER model to the multilingual XLM-T model in order to study cross-cultural differences. Therefore, to calibrate the results obtained with XLM-T with respect to those obtained with VADER, both models were applied to the English tweets in the baseline dataset and the numbers of negative relationships in the full and active Ego Networks were computed. The results, as well as the differences in negativity between the full and active networks for both models, are shown in Table~\ref{tab:model_comparison}. When these two sets of negativities are compared, XLM-T appears to sign relationships between 8 and 10 percentage points more negative, although the difference between the full and active networks is almost the same. This drift between VADER and XLM-T should be kept in mind when comparing the results obtained with the multilingual classifier against the results in the initial SENM paper~\cite{Tacchi_2022}, obtained using VADER.

\begin{table}
  \caption{Comparison of the percentages of negative relationships of VADER and XLM-T in the full and active Ego Networks of the baseline dataset}
  \label{tab:model_comparison}
  \begin{tabular}{l r r r r}
    \toprule
    Model & Full & & Active & Difference\\
    \midrule
    VADER & 24.05\% & & 40.31\% & +16.26\\
    XLM-T & 32.25\% & & 49.90\% & +17.65\\
  \end{tabular}
\end{table}

\subsection{Full and Active Ego Networks}
\label{subsec:full_active_ego_networks}
The first analysis was carried out after obtaining the full and active networks, via the preprocessing stages described in Subsection~\ref{subsec:preprocessing}, as well as the relationship signs, mentioned in Section~\ref{subsec:computation_of_signed_ego_networks}. Recall that active Ego Networks include only relationships that involve regular cognitive effort on the Ego's part, while full Ego Networks include all social interactions the Ego has been involved with. The percentages of negative relationships in each dataset's full and active networks can be seen in Table~\ref{tab:full_vs_active_negatives}.

\begin{table}
  \caption{Percentages of negative relationships in the full and active Ego networks}
  \label{tab:full_vs_active_negatives}
  \begin{tabular}{l r r r r}
    \toprule
    Dataset & Full & & Active & Difference\\
    \midrule
    Baseline & 32.25\% & & 49.90\% & +17.65\\
    Mediterranean & 45.76\% & & 60.08\% & +14.32\\
    Northern Europe & 38.31\% & & 54.66\% & +16.34\\
    West Africa & 32.77\% & & 50.29\% & +17.52\\ 
    South America & 42.83\% & & 65.67\% & +22.85\\
    \bottomrule
    Italian Journalists & 40.61\% & & 63.87\% & +23.26\\
    Dutch Journalists & 34.55\% & & 57.65\% & +23.10\\
    Brazilian Journalists & 44.80\% & & 64.93\% & +20.13\\
    Italian Reality TV & 52.13\% & & 64.97\% & +12.84\\
    Dutch Reality TV & 46.39\% & & 68.35\% & +21.97\\
    Brazilian Reality TV & 44.18\% & & 69.47\% & +25.29\\
    \bottomrule
  \end{tabular}
  \vspace{-10pt}
\end{table}

Looking at the levels of negativity across the networks, one can see that the generic users have negativities between 32.25\% and 45.76\% for the full networks and between 49.90\% and 65.67\% for the active. By contrast, the themed users are generally more negative, ranging between 34.55\% and 52.13\% for the full networks and between 57.65\% and 69.47\% for the active. When XLM-T's shift towards negativity is accounted for, these results are very similar to those of the previous work. Those previous ranges being 16.45\% to 24.22\%, for full networks, and 21.83\% to 40.31\%, for active networks, for generic users and 27.15\% to 31.58\%, for the full networks, and 45.23\% to 54.89\%, for active networks, for themed users~\cite{Tacchi_2022}. Two datasets, South America and Dutch Journalists don't quite seem to fit their groups at first glance, being slightly too negative and too positive respectively. However, the Dutch Journalists are still more negative than their generic Northern Europe equivalent for the active network, and so this difference could be due to cultural differences in journalism between the Dutch and other journalist datasets. Similarly, the conspicuously high negativity of South America could be caused by the types of topics those users are engaged with, this is discussed in more detail in Subsection~\ref{subsec:most_popular_topics}.

Next, comparing the negative percentages between the full and active networks, one can see that, as expected, all of the datasets are more negative in the active network. What's more, these increases in negativity are generally greater for thematic users than for generic users (apart from the notable exception of Italian Reality TV, which was already very negative in the full network). Both these findings mirror previous works~\cite{Toprak_2021,Tacchi_2022}, which have hypothesised that users who are more engaged with the platform exhibit more negativity. 

Finally, there appear to be some cultural differences, which are reflected by the largely consistent order of the negativities; with the Dutch datasets being the least negative, followed by the Italian and then the Brazilian. This order is observable in both the full and active networks of the journalists as well as in the active network of the generic users. By contrast, these cultural differences are not observable for the Reality TV datasets. Moreover, the Reality TV group appear to be generally more negative with regard to their generic counterparts. They are fairly similar regardless of culture, especially for the active networks, where there are only 4.5 percentage points between the most and the least negative datasets (compared to 6.22 for journalists and 11.01 for the 3 corresponding regional datasets). This hints that the common interest of these individuals (in this case reality TV) is overpowering their cultural differences.

\subsection{Unsigned Ego Network Analysis}
\label{subsec:ego_network_analysis} 
Next, the structure of the Ego Networks is analysed, before considering the signs of the relationships. This entails computing the optimum number of social circles as well as the sizes of those circles, for each Ego. Due to slight differences between individuals, the optimum number of circles can vary between Egos. This is a regular observation in Ego Networks research~\cite{Arnaboldi_2013a,Dunbar_2015} and it is standard practice to focus on Egos with a common number of optimum circles when conducting any analyses. This number is usually 5, and as this is also the closest whole number to the average optimum circles in 5 out of the 9 datasets, with the other 4 also being close to 5. The mean optimum circle numbers and the mean network sizes of the corresponding Egos can be seen in Table~\ref{tab:optimum_circle_and_mean_network_size}. Therefore, all following results in this paper include only Egos whose optimum circle is 5, unless otherwise stated.

\begin{table}
  \caption{Mean number of optimum circles [95\% confidence intervals] and number of Egos with 5 circles}
  \label{tab:optimum_circle_and_mean_network_size}
  \begin{tabular}{lcr}
    \toprule
    Dataset & Mean \# Circles & Egos\\
    \midrule
    Baseline & 5.50 [5.46, 5.54] & 1,160\\
    Mediterranean & 5.62 [5.53, 5.71] & 227\\
    Northern Europe & 5.68 [5.56, 5.79] & 162\\
    West Africa & 5.94 [5.80, 6.09] & 84\\
    South America & 5.36 [5.19, 5.54] & 59\\
    \bottomrule
    Italian Journalists & 5.72 [5.52, 5.93] & 51\\
    Dutch Journalists & 5.45 [5.38, 5.51] & 440\\
    Brazilian Journalists & 5.46 [5.27, 5.65] & 50\\
    Italian Reality TV & 5.48 [5.25, 5.71] & 40\\
    Dutch Reality TV & 5.47 [5.29, 5.66] & 62\\
    Brazilian Reality TV & 5.25 [4.95, 5.55] & 48\\
    \bottomrule
  \end{tabular}
\end{table}

The mean numbers of Alters at each level of the Ego Networks can be seen in Table~\ref{tab:circle_sizes}. These numbers are very close to the expected sizes: 1.5, 5, 15, 45-50, 150~\cite{Dunbar_2015}. As is often found, especially in online data~\cite{Arnaboldi_2017}, the outer circles are lower than expected, but this is not overly surprising as even the most active Twitter user is unlikely to develop the entirety of their Ego Network on the platform. Therefore, there appear to be no abnormalities in the unsigned Ego Networks of the included datasets.

\begin{table}
  \caption{Mean circle sizes of Egos with an optimum circle number of 5}
  \label{tab:circle_sizes}
  \begin{tabular}{l c c c c c}
    \toprule
    Dataset & C\textsubscript{1} & C\textsubscript{2} & C\textsubscript{3} & C\textsubscript{4} & C\textsubscript{5}\\
    \midrule
    Baseline & 1.78 & 6.16 & 16.86 & 44.19 & 125.91\\
    Mediterranean & 1.70 & 5.60 & 14.67 & 38.83 & 120.41\\
    Northern Europe & 1.80 & 5.88 & 17.12 & 45.34 & 131.12\\
    West Africa & 1.65 & 5.60 & 15.64 & 39.71 & 118.81\\ 
    South America & 1.80 & 5.76 & 15.71 & 39.92 & 118.29\\
    \bottomrule
    Italian Journalists & 1.12 & 3.57 & 10.59 & 33.14 & 120.10\\
    Dutch Journalists & 1.66 & 5.51 & 15.54 & 43.08 & 122.69\\
    Brazilian Journalists & 1.78 & 5.90 & 15.66 & 41.62 & 116.48\\
    Italian Reality TV & 1.63 & 5.15 & 13.85 & 35.60 & 103.65\\
    Dutch Reality TV & 1.61 & 5.29 & 14.29 & 37.16 & 98.63\\
    Brazilian Reality TV & 1.58 & 4.65 & 12.08 & 31.29 & 96.63\\
    \bottomrule
  \end{tabular}
  \vspace{-11pt}
\end{table}

\subsection{Signed Ego Network Analysis}
Combining the Ego Networks with the signed relationships as described in Subsection~\ref{subsec:computation_of_signed_ego_networks}, it is possible to observe the number of negative relationships at each level of the Ego Networks. The means and percentages of negative relationships per circle can be viewed in Table~\ref{tab:circle_negativities}. Since circle sizes vary significantly, percentages are more informative than absolute numbers. As previously mentioned in Subsection~\ref{subsec:ego_network_analysis}, circle 5 tends to not be entirely complete when data comes from an online source, therefore this analysis will focus mainly on circles 1 to 4.

\begin{table*}
  \caption{Mean number of negative relationships for Egos with 5 circles. The percentages of negative relationships for each circle size and total active network are displayed in parentheses, with the most negative circle of each dataset in bold.}
  \label{tab:circle_negativities}
  \begin{tabular}{l c c c c c c}
    \toprule
    Dataset & C\textsubscript{1} & C\textsubscript{2} & C\textsubscript{3} & C\textsubscript{4} & C\textsubscript{5} & Active\\
    \midrule
    Baseline & 1.00 (56.25\%) & \textbf{3.63 (58.84\%)} & 9.72 (57.64\%) & 24.28 (54.95\%) & 63.71 (50.60\%) & 49.90\%\\
    Mediterranean & \textbf{1.25 (73.58\%)} & 4.06 (72.54\%) & 10.38 (70.77\%) & 27.07 (69.70\%) & 76.85 (63.82\%) & 60.08\%\\
    Northern Europe & \textbf{1.26 (69.86\%)} & 3.94 (67.05\%) & 11.04 (64.48\%) & 27.45 (60.54\%) & 70.67 (53.89\%) & 54.66\%\\
    West Africa & 0.92 (55.40\%) & \textbf{3.18 (56.81\%)} & 8.75 (55.94\%) & 21.25 (53.51\%) & 60.80 (51.17\%) & 50.29\%\\
    South America & 1.37 (76.42\%) & \textbf{4.42 (76.76\%)} & 12.00 (76.38\%) & 28.71 (71.93\%) & 75.03 (63.43\%) & 65.67\%\\
    \bottomrule
    Italian Journalists & \textbf{1.00 (89.47\%)} & 3.12 (87.36\%) & 8.67 (81.85\%) & 25.80 (77.87\%) & 84.80 (70.61\%) & 63.87\%\\
    Dutch Journalists & \textbf{1.19 (71.60\%)} & 3.90 (70.74\%) & 10.78 (69.34\%) & 28.13 (65.30\%) & 71.50 (58.27\%) & 57.65\%\\
    Brazilian Journalists & 1.14 (64.04\%) & 4.42 (74.92\%) & \textbf{11.94 (76.25\%)} & 30.20 (72.56\%) & 77.02 (66.12\%) & 64.93\%\\
    Italian Reality TV & 1.08 (66.15\%) & 3.85 (74.76\%) & \textbf{10.58 (76.35\%)} & 26.38 (74.09\%) & 71.38 (68.86\%) & 64.35\%\\
    Dutch Reality TV & 1.15 (71.00\%) & 3.97 (75.00\%) & \textbf{10.97 (76.75\%)} & 27.73 (74.61\%) & 67.42 (68.36\%) & 68.35\%\\
    Brazilian Reality TV & \textbf{1.31 (82.89\%)} & 3.83 (82.51\%) & 9.92 (82.07\%) & 24.31 (77.70\%) & 67.73 (70.09\%) & 69.47\%\\
    \bottomrule
  \end{tabular}
\end{table*}

First, the highest percentages of negative relationships appear towards the inner circles. Specifically, C\textsubscript{1} is the most negative circle for 5 of the datasets (Mediterranean, Northern Europe, Italian Journalists, Dutch Journalists and Brazilian Reality TV), with C\textsubscript{2} and C\textsubscript{3} also holding the highest negativities for 3 datasets each (those being Baseline, West Africa and South America for the former and Brazilian Journalists, Italian Reality TV and Dutch Reality TV for the latter). Recall that the innermost circles (namely C\textsubscript{1} and C\textsubscript{2}) contain the most intimate and engaged social relationships of the Egos. Hence, a high percentage of negative relationships in those circles would suggest a social life dominated by negative connections. While these findings do broadly match past observations of similar datasets, the number of C\textsubscript{1}s displaying the highest negativity is lower than might be expected. This lack of negative relationships at the innermost layers could occur due to stronger effects of social capital, which would make the inner circles "artificially" more positive. Indeed, this is likely to occur in datasets that contain close-knit communities where there is more social pressure to get along (or at least to give the appearance of getting along). It could be pertinent to further investigate this in future research.

Next, observing how the percentages vary across the circles, it appears that the generic users and journalists are more or less in line with expectations. Indeed, the differences between the most and least negative circles (C\textsubscript{1}-C\textsubscript{4}) are, in percentage points and in the order the datasets were initially presented, 3.89, 3.88, 9.32, 2.43 and 4.83 for the generics and 11.60, 6.30 and 12.21 for the journalists. However, there is some overlap between these groups; with Northern Europe being less variable than its Dutch Journalist counterpart. The reality TV datasets appear in between these two groups with differences of 10.20, 5.75 and 5.19 percentage points respectively, which could correspond to a middling level of engagement with the platform.

Interestingly, some of the datasets that appeared to be very similar when observing their active network percentages, do not appear so similar when looking at their circles. For example, the active networks of all the TV datasets are relatively close, however, the Brazilians' circles differ starkly from those of the other two. Further analysis focused on these aspects could provide interesting and more detailed insights.




\subsection{Most Popular Topics}
\label{subsec:most_popular_topics}

As the theme of discussion seems to be correlated with the proportion of negative relationships, this interaction was investigated at a more granular level through a topic-based analysis. Specifically, the top 20 hashtags and the top 20 words were examined for both the full and active networks, for all Egos (not just those with an optimum circle number of 5)\footnote{Full lists of these words and hashtags are available in the Appendix of the technical report, which can be found at: \url{https://anonymous.4open.science/r/cultural-senm-technical-report-6674/Cultural_Differences_in_Signed_Ego_Networks_on_Twitter_An_Investigatory_Analysis_Technical_Report.pdf}}. Both the hashtags and the words were standardised, removing diacritics, punctuation and capitalisation. Words were also removed if they were stopwords or if they were equal to or less than 4 letters in length, these latter 2 restrictions were not imposed on the hashtags.

As hashtags are generally used to indicate the topics within a tweet, these were the main focus of this analysis. The hashtags in the active networks were manually assigned 1 of 6 labels, corresponding to different topics. These labels were: "political" for politicians, governments, political topics or politic-only news channels, "COVID" for anything relating to the COVID-19 pandemic, "climate" for topics relating to green energy, renewable or the climate, "religious" for religious topics, "news" for general news services (i.e. not specifically political or religious) and "general" for everything else (these hashtags and their labels can be seen in Appendix~\ref{app:most_popular_topics}). Using the percentage of negative relationships in the active networks, correlations were then calculated using Pearson's R for the number of times each topic appeared in the top 20, as well as the total and proportional\footnote{The proportional number being the how many times all the hashtags of each topic were mentioned in the dataset, divided by the total number of mentions of all hashtags in the top 20 for that dataset.} number of times those hashtags were mentioned. For this test, there were 9 degrees of freedom. The only variable that had a significant correlation with negativity was the number of "general" topics in the top 20, r(9) = -.64, p = .035, meaning that there is a negative correlation between the number of "general" topics users frequently discuss and the number of negative relationships they have. One potential reason for this, given that individuals are more likely to engage in negative exchanges with someone whose beliefs differ from their own~\cite{Hutchens_2015}, is that higher levels of negativity are more likely to occur when the topic of conversation is specific, as strong opinions are more likely to have formed, as opposed to non-specific topics, where strong opinions are less likely to have formed. Indeed, this has previously been found for topics which have clear "sides", such as in politics~\cite{Coe_2014}. On top of this, being online means that you are more likely to be exposed to a greater range of opinions~\cite{Heatherly_2017} and those who spend more time on the platform (i.e. those who are the most engaged) are more likely to find someone they disagree with, thereby having an increased likelihood of negative exchanges. This could explain the higher observed negativities for journalists and TV watchers.

As mentioned in Subsection~\ref{subsec:full_active_ego_networks}, South America is surprisingly negative for a generic user dataset, even displaying slightly more negativity than the Brazilian Journalists. However, the top 3 hashtags in the active network of this dataset, as well as 8 out of the top 20, are news-related. All the other generic datasets have exactly 1 hashtag in the news category. This suggests that users in South America may be using Twitter more as a news site than as a social media platform, which would explain why the generic users from this region are so similar to the journalists.

While the results of the analysis on the top 20 words revealed information that was very similar to that of the hashtags, one additional and rather interesting finding did arise: the usage of the word "Allah" in the West Africa dataset. There are many common phrases used in Muslim cultures that contain the word "Allah"; for example, "may Allah grant you health" or "may Allah strengthen the noble"~\cite{Rababah_2012}. Although there are differences in how these types of phrases are used between Muslin countries, they usually contain positive sentiment that is much stronger than the equivalent phrases used by other cultures; such as "hello" or "good morning". This could go some way towards explaining why West Africa is the most positive regional dataset and the second most positive overall. What's more, greetings in some West African cultures are known to be highly formalised and predominantly phatic (i.e. with the aim of establishing or maintaining social relationships)~\cite{Goody_1972}. Thus, there could be a double effect towards positive communications. This specific use of certain phrases is a cultural difference that is relatively easy to detect when analysing social media data and could be directly affecting how individuals communicate, potentially enabling some very interesting insights about cultural differences to be revealed in future work.

\section{Conclusion}
\vspace{-2pt}
This paper has taken the recent concept of a signed Ego Network and observed it using a culturally diverse cross-section of Twitter data from across the globe. This included 11 datasets (8 of which were collected as part of this work), split into 3 different categories of user: generic, journalist and reality TV watchers. Notably, these datasets contain multilingual data, which is far more challenging to deal with than the English-only data of the initial SENM work~\cite{Tacchi_2022}. First, previous SENM results were replicated using the new datasets. The percentages of negative relationships in both the full and active networks, the increase in negativity from full to active network, the sizes of the Ego circle, and the distribution of negativity across the circles of generic users and journalists were all in line with expectations from previous work. 

Next, these replicated results were complimented by 3 key, novel observations: 
(i) there appear to be some minor differences in negativity due to culture in the generic datasets but these seem to get overwhelmed when the network is centred around a specific topic; 
(ii) networks that are "themed" around a certain topic are consistently more negative than generic networks, both overall and at each level of the SENM; 
(iii) "general" is the only topic to have a significant correlation with the percentage of negativity relationships, further supporting the idea that Twitter conversations based on a specific topic are more likely to lead to negativity.

In addition to these main findings, some limitations regarding the current work should be considered. For one, it is extremely difficult to guarantee a standard level of performance when conducting the task of sentiment analysis on a corpus of data in multiple languages. Indeed, while as much effort as possible was made to ensure the standardisation of the performances across all the languages used in this paper, the results for some of the datasets may not be as reliable as for others. This could be caused, for example, by a bias in XLM-T training data that is present for one of the languages but not for the others. Next, although the datasets used in this paper display some cultural effects that are sometimes being overpowered by the topics of conversation, this represents a somewhat small sample size. So, while the observations made in this paper may provide an important initial glimpse at some interesting cultural differences, it would be pragmatic for future research to replicate these findings with even more datasets, to further illuminate these observations.

Finally, some other avenues for future work have been suggested. For instance, the interplay between the "closeness" of a community and the percentage of negative relationships it has could be investigated within the context of the SENM, especially given the seemingly paradoxical effects of social capital and the more negative inner circles. Similarly, a deep dive into the usage of certain types of language use (e.g. greetings, compliments, etc.) could be observed at different circles within the SENM. This could be done not only between cultures but also between online communities that have grown around specific topics. On a similar line of reasoning, it could potentially prove fruitful to gather more datasets themed on different topics, such as sports, tech, etc., to see if any further patterns emerge regarding the effect of topic on negativity.

\vspace{-5pt}
\begin{acks}
\vspace{-4pt}
This work was partially supported by the H2020 HumaneAI-Net (952026) and SoBigData++ (871042) projects, by the CHIST-ERA-19-XAI010 SAI project. C. Boldrini, A. Passarella and M. Conti's work was partly funded by the PNRR - M4C2 - Investimento 1.3, Partenariato Esteso PE00000013 - "FAIR", funded by the European Commission under the NextGeneration EU programme. C. Boldrini's work was also partially funded by the PNRR - M4C2 - Investimento 1.4, Centro Nazionale CN00000013 - "ICSC", funded by the European Commission under the NextGeneration EU programme.
\end{acks}

\bibliographystyle{ACM-Reference-Format}
\bibliography{references}

\newpage
\appendix

\section{Baseline User Locations}
\label{app:locations}

\subsection{Countries}
\label{subsec:countries}
The locations of engaged users in the baseline dataset were gathered by passing their provided location into the Google Maps API. The "UNK" designation was assigned to users for whom it was not possible to obtain a location. The number of Egos and relationships for each location are displayed in Table~\ref{tab:locations_country}, only countries with more than 2 Egos are included.

\begin{table}
  \caption{Number of Egos and relationships for each country in the baseline dataset}
  \label{tab:locations_country}
  \begin{tabular}{lrr}
    \toprule
    Country & Egos & Relationships\\
    \midrule
    United States & 899 & 84,018\\
    UNK & 797 & 76,011\\
    United Kingdom & 129 & 12,850\\
    Nigeria & 61 & 6,247\\
    Canada & 60 & 4,691\\
    Australia & 17 & 1,897\\
    Spain & 15 & 1,223\\
    Brazil & 13 & 961\\
    South Africa & 10 & 1,017\\
    Ireland & 8 & 892\\
    India & 8 & 868\\
    Netherlands & 8 & 802\\
    Mexico & 6 & 363\\
    Sweden & 6 & 532\\
    Italy & 6 & 651\\
    Germany & 5 & 410\\
    France & 5 & 534\\
    Belgium & 4 & 279\\
    Colombia & 4 & 424\\
    Jamaica & 4 & 272\\
    Jordan & 3 & 191\\
    Japan & 3 & 315\\
    Poland & 3 & 296\\
    \bottomrule
  \end{tabular}
\end{table}

\subsection{Continents}
Next, the countries from Subsection~\ref{subsec:countries} were compiled into continents, except that all locations were included regardless of the number of Egos. These results are displayed in Table~\ref{tab:locations_continents}.

\begin{table}
  \caption{Number of Egos and relationships for each continent in the baseline dataset}
  \label{tab:locations_continents}
  \begin{tabular}{lrr}
    \toprule
    Continent & Egos & Relationships\\
    \midrule
    North America & 971 & 89,552\\
    UNK & 799 & 76,240\\
    Europe & 207 & 19,765\\
    Africa & 80 & 8,147\\
    Asia & 30 & 2,827\\
    South America & 23 & 2,024\\
    Oceania & 19 & 1,930\\
    \bottomrule
  \end{tabular}
\end{table}

\section{Most Popular Topics}
\label{app:most_popular_topics}




\subsection{Top 20 Active Hashtags with Topics}
\label{app:20_hashtags_with_topics}
\begin{figure*}[h]
  \centering
  \includegraphics[scale=0.57]{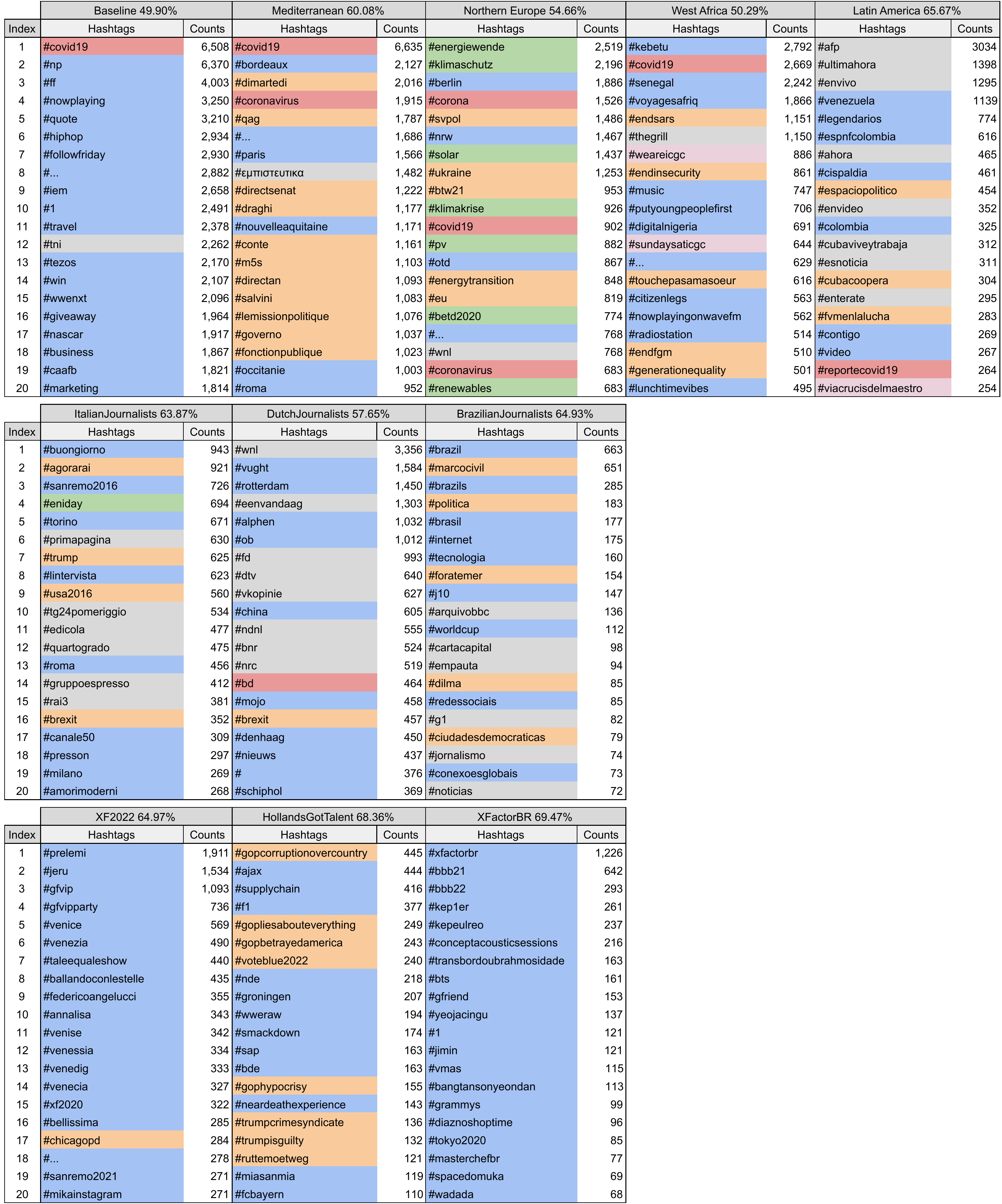}
  \caption{The 20 most used hashtags in the active networks, colour-coded by topic: "Political" in orange, "General" in blue, "COVID" in red, "Climate" in green, "Religious" in pink and "News" in grey. The percentages of negative relationships in the active networks of each dataset are also given alongside the names.}
  \Description{The top 20 most used hashtags (and their counts) for each dataset in the active networks of each of the 11 datasets, colour-coded by topic ("Political" in orange, "General" in blue, "COVID" in red, "Climate" in green, "Religious" in pink and "News" in grey). Unfortunately, due to the amount of data represented in this figure, it is not possible to list all the hashtags within the 2,000-character limit.}
  \label{fig:top_20_hashtags_coloured}
\end{figure*}

The top 20 most used hashtags for active networks can be seen in Figure~\ref{fig:top_20_hashtags_coloured}. They are colour-coded by topic ("Political" in orange, "General" in blue, "COVID" in red, "Climate" in green, "Religious" in pink and "News" in grey), as described in Subsection~\ref{subsec:most_popular_topics}.

\end{document}